\documentclass[entropy,article,submit,pdftex,moreauthors]
{Definitions/mdpi} 
\renewcommand{\linenumbers}{}
\usepackage{float}
\firstpage{1} 
\pubvolume{1}
\issuenum{1}
\articlenumber{0}
\pubyear{2023}
\copyrightyear{2023}
\datereceived{ } 
\daterevised{ } 
\dateaccepted{ } 
\datepublished{ } 
\hreflink{https://doi.org/} 

\Title{Linking QKD testbeds across Europe }

\TitleCitation{Title}

\Author{Max Brauer$^{1}$,
Rafael J.\ Vicente$^{2}$,
Jaime S. Buruaga$^{2}$,
Rubén B.\ Méndez$^{2}$,
Ralf-Peter Braun$^{1}$,
Marc Geitz$^{1}$,
Piotr Rydlichkowsk$^{3}$,
Hans H.\ Brunner$^{4}$,
Fred Fung$^{4}$,
Momtchil Peev$^{4}$,
Antonio Pastor$^{5}$,
Diego Lopez$^{5}$,
Vicente Martin$^{2}$,
and Juan P.\ Brito$^{2}$}

\AuthorNames{Firstname Lastname, Firstname Lastname and Firstname Lastname}

\AuthorCitation{Lastname, F.; Lastname, F.; Lastname, F.}

\address{%
$^{1}$ \quad Deutsche Telekom AG, T-Labs, Berlin, Germany\\
$^{2}$ \quad DLSIIS and Center for Computational Simulation, Universidad Politécnica de Madrid, Madrid, Spain.\\
$^{3}$ \quad Poznan Supercomputing and Networking Center, Poznań, Poland\\
$^{4}$ \quad Munich Research Center, Huawei Technologies Duesseldorf GmbH, Munich, Germany\\
$^{5}$ \quad Telefónica gCTIO/I+D, Madrid, Spain\\
}

\corres{Correspondence: juanpedro.brito@upm.es; Tel.: +34-910673073 (J.P.B.); marc.geitz@telekom.de (M.G.)}

\secondnote{These authors contributed equally to this work.}

\usepackage[colorinlistoftodos]{todonotes} 
\usepackage{ulem}
\usepackage{makecell}

\abstract{Quantum-key-distribution (QKD) networks are gaining importance and it has become necessary to analyze the most appropriate methods for their long-distance interconnection.  In this paper, four different methods of interconnecting remote QKD networks are proposed.  The methods are used to link three different QKD testbeds in Europe, located in Berlin, Madrid, and Poznan.  Although long-distance QKD links are only emulated, the used methods can serve as a blueprint for a secure interconnection of distant QKD networks in the future.  Specifically, the presented approaches combine, in a transparent way, different fiber and satellite physical media, as well as common standards of key-delivery interfaces.  The testbed interconnections are designed to increase the security by utilizing multipath techniques and multiple hybridizations of QKD and post quantum cryptography (PQC) algorithms.}


\newcolumntype{Y}{>{\centering\arraybackslash}X}

\begin{document}



\section{Introduction}

The Shor's algorithm \cite{Shor} for quantum computers can solve the fundamental mathematical problems, on which currently used public-key cryptography (PKC) is based, in polynomial time. Therefore, the security PKC offers will be lost when mature quantum computers are available.  It should be noted that the security of traditional PKC is based on a complexity assumption: the fundamental problems enabling it, such as factoring of large integers, the discrete-logarithm problem, and the elliptic-curve discrete-logarithm problem (rooted itself in the algebraic structure of elliptic curves over finite fields), are not solvable in polynomial time, which makes performing some inversion tasks intractable.  This assumption of ``non-polynomiality'' has been demonstrated to be false for quantum computers, specifically by the mentioned Shor algorithm. 

A presently developed ``remedy'' technology is post-quantum cryptography (PQC) \cite{Nist}, which is a quantum-safe version of PKC. PQC is a broad concept but at its core lie public key algorithms that similar to their "classical" counterparts are based on mathematical problems that cannot be inverted in polynomial time by any known algorithm including Shor's one. In this sense they are "Shor resistant" and assumed to be intractable also for any other quantum algorithms.  The mentioned assumptions are underpinned by mathematical-complexity arguments in their versions for quantum computing.  The specific status of PQC development is represented by three protocols in draft standardization phase, belonging to two classes: key agreement ones based on quantum-safe public key encryption-decryption of secret key - Key Encapsulation Mechanisms (KEM), in what follows PQC KEM, and  digital Signature ones (SIG), in what follows PQC SIG \cite{NIST_CSRC}.

Quantum key distribution (QKD) is still another technology that provides symmetric keys to two distant parties but employs protocols which are information-theoretically secure (ITS), the utmost level of security.  These protocols are based on fundamental features of quantum mechanics and are not vulnerable against any type of adversaries, irrespective of their resources.  This includes also eavesdroppers equipped with quantum computers.  

Present QKD protocols and implementations have only a finite reach.  This reach can be extended by using chains or networks of trusted nodes, e.g., trusted satellites.  These reach extensions follow protocols that are also ITS, as is the original QKD, albeit assuming the enforcement of additional non-cryptographic (organizational) measures, which allow absolute trust in the integrity of intermediate nodes.  In the following these extension protocols are called ``key forwarding''.  Such extensions are intensively tested in the European Union and China \cite{Berlin-testbed,Poznan-testbed,martin2023madqci,QKD_SDN,China_Network}. 

Like any other technology, QKD, QKD-network, and PQC implementations are security-wise weaker than the respective ideal protocols. The implementations are vulnerable to so called ``side channels'' that generally represent the differences between implemented and ideal protocols.  QKD protocol security is a mathematical theorem, which follows from several (sufficient) conditions that cannot realistically be guaranteed in QKD implementations.  The discussion of such issues is beyond our present objective and is generally addressed by the procedures of certification of security technologies \cite{ETSI-016,ISO23837}.  In this text, it is acknowledged that an implementation is always less secure than an underlying protocol by using the term ``side channels'' to denote all differences.

The goal of this paper is to demonstrate a feasible connection of present-day European metropolitan-area QKD networks, as developed in OpenQKD and planned in EuroQCI \cite{Openqkd,Euroqci}, before long-distance QKD is available and chains of trusted nodes, quantum repeater links, or QKD satellites are installed in Europe. In this work, the long-distance QKD connections are only emulated QKD links. 

Independently, all links were realized as dual-technology key generation, based on (emulated) QKD and PQC. This is the maximum that can be demonstrated without true long distance QKD. We therefore present a blueprint of future design featuring higher security and simulate it by emulated QKD links where true ones are not yet technically available. This shows a potential utilization of two methods in parallel \cite{Muckle} and emulates a ``crypto-agility'' realization, the secure combination or hybridization of different security technologies.  Such a principal strategy is generally reasonable in a dynamic field, in which the security of protocols can be quickly re-evaluated.  Specifically, the combination of multiple security-protocol implementations, that might be vulnerable to yet unknown but most likely different side channels, was demonstrated in this work.  Such a scheme reduces risks, especially against non-coinciding side channels that might originally jeopardize the implementation of a single protocol.

It must be emphasized that only the parallel combination of QKD and PQC can increase security, while the sequential combination of protocols can only lead to a security reduction. In the case of a parallel implementation a potential attacker must attack all vulnerabilities independently (except for their potential common intersection). In the case of a sequential combination, it is sufficient to attack the weakest element from the perspective of the attacker. For a more detailed clarification some combination possibilities and the resulting security level are listed in tables~\ref{tab01} and \ref{tab02}.  The benefit of parallel combinations can be found in table~\ref{tab01}, while table~\ref{tab02} shows the issues of serial combinations.  The fourth row in table~\ref{tab01} is introduced, because it is often seen as convenient for managing the otherwise tedious authentication issues in large and evolving QKD networks.  QKD will only be ITS, if the keys required for message integrity (message-authentication) in  post-processing are generated using ITS mechanisms, e.g., with pre-shared keys (PSK) for the first QKD round and using a portion of the QKD generated key subsequently\cite{MQuade,Pacher}.  Using PQC for authentication limits the security level of the key generation to that of this authentication method and the ITS advantage of QKD is lost.  In that case one could use PQC directly for key generation.  In the end, the adopted level of security must be a conscious decision of the network operator.

\begin{table}[h]
\caption{Security level of various combinations of different key-exchange technologies in parallel in a symbolic set-theoretic representation.  This could be, e.g., discrete variable (DV) and continuous variable (CV) QKD protocols in pure QKD networks, PQC KEM and PQC SIG algorithms in pure PQC networks and combinations of pairs of technologies for different tasks in mixed networks.  
QKD implementations are ITS minus ($\setminus$) side channels (SCs), while the security level of PQC is based on mathematical-complexity (MC) minus SCs.  The combined security level of parallel key generation is the union ($\cup$) of the individual security levels.  For key exchanges of the same security level this simplifies to this level but minus the intersection ($\cap$) of the SCs. If not otherwise noted, QKD with PSK will be assumed. Note that this classification is independent on whether the different technologies are applied over the same or different network paths.\label{tab01}}
\begin{tabularx}{\textwidth}{| c | Yc | Yc | c |}
\toprule
\bf ID
& \bf Key exchange 1
& \bf Security
& \bf Key exchange 2
& \bf Security
& \bf Combined security level
\\ \midrule
1
& QKD$_1$
& ITS $\setminus$ SCs$_1$
& QKD$_2$
& ITS $\setminus$ SCs$_2$
& ITS $\setminus \{\mathrm{SCs}_1 \cap \mathrm{SCs}_2\}$
\\ \midrule
2
& PQC$_1$
& MC $\setminus$ SCs$_1$
& PQC$_2$
& MC $\setminus$ SCs$_2$
& MC $\setminus \{\mathrm{SCs}_1 \cap \mathrm{SCs}_2\}$
\\ \midrule
3
& QKD$_1$
& ITS $\setminus$ SCs$_1$
& PQC$_2$
& MC $\setminus$ SCs$_2$
& $\left(\mathrm{ITS} \setminus \mathrm{SCs}_1\right) \cup \left(\mathrm{MC} \setminus \mathrm{SCs}_2\right)\}$
\\ \midrule
4
& QKD$_1$ + PQC SIG
& MC $\setminus$ SCs$_1$
& QKD$_2$ + PQC SIG
& MC $\setminus$ SCs$_2$
& MC $\setminus \{\mathrm{SCs}_1 \cap \mathrm{SCs}_2\}$
\\ \bottomrule
\end{tabularx}
\end{table}

\begin{table}
\caption{Security level of various combinations of different key-exchange technologies in series in a symbolic set-theoretic representation.  The combined security level of serial key generation is the intersection of the individual security levels.  For the examples in this table this is equal to the respectively lower security level minus the union of the SCs.\label{tab02}}
\begin{tabularx}{\textwidth}{| c | Yc | Yc | c |}
\toprule
\bf ID
& \bf Key exchange 1
& \bf Security
& \bf Key exchange 2
& \bf Security
& \bf Combined security
\\ \midrule
1
& PQC$_1$
& MC $\setminus$ SCs$_1$
& QKD$_2$
& ITS $\setminus$ SCs$_2$
& MC $\setminus \{\mathrm{SCs}_1 \cup \mathrm{SCs}_2\}$
\\ \midrule
2
& QKD$_1$
& ITS $\setminus$ SCs$_1$
& QKD$_2$
& ITS $\setminus$ SCs$_2$
& ITS $\setminus \{\mathrm{SCs}_1 \cup \mathrm{SCs}_2\}$
\\ \bottomrule
\end{tabularx}
\end{table}

At the moment, the cryptographically most secure combination is a direct, any-to-any PQC key exchange combined with an any-to-any QKD key exchange relying on key forwarding.  For this paper, the PQC links were established directly between any two nodes, irrespective of metro-network boundaries.  In contrast, the QKD key generation between nodes belonging to different metropolitan-area QKD networks used key forwarding over emulated, long-distance QKD connections only between selected border nodes.

Additionally, a two-path approach, i.e., one that utilizes two different paths, will reduce risks even if a single implementation of a single technology is used.  Risks will be further reduced, if different implementations and moreover different technologies, deployed over different paths, are used.  In the following these approaches will be dubbed as ``two-factor'' ones, whereby in reality multiple factors are involved.  Actual connections can be complex combinations of different paths and the different options listed in table~\ref{tab01} and \ref{tab02}.  In this work, most PQC links were realized as a two-technology and two-path combination over a ground link and a satellite link, while the QKD key forwarding retrieved keys from QKD modules of different vendors in series but also in parallel. 

\begin{figure}[H]
\centering
\includegraphics[width=\textwidth]{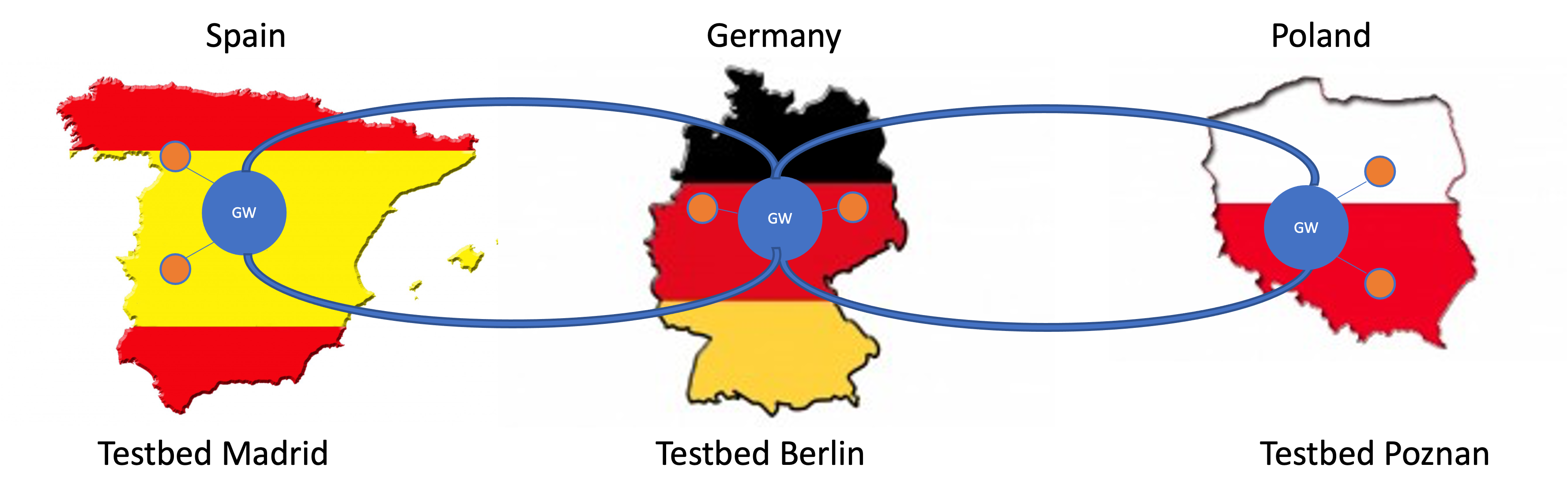}
\caption{Connection of the quantum testbeds in Madrid (left), Berlin (middle), and Poznan (right) with emulated, long-distance QKD links.  The key exchange is indicated by the curved blue lines, which connect dedicated QKD gateway nodes (blue circles) in each testbed.  The other QKD nodes in the respective testbeds are indicated by orange circles.\label{fig04}}
\end{figure}

The QKD-network testbeds used for the demonstration are located in Berlin (Germany), Madrid (Spain) and Poznan (Poland).  These are symbolically shown in Figure~\ref{fig04}. Each testbed represents a metropolitan-area quantum-optical network. The distances Berlin-Madrid and Poznan-Berlin are 1860 km and 230 km, respectively.  The testbeds are too far apart for state-of-the-art, point-to-point QKD links, which is an additional motivation for the approach outlined in this Section (see option 1 in table~\ref{tab02}).

The long-distance QKD links in this work have been emulated using PQC KEM crypto protocols, as discussed in detail below. For transmission purposes either normal (classical) terrestrial communication or a combination of classical terrestrial and classical satellite communication has been used.

\section{Participating QKD testbeds}

\subsection{QKD testbed in Berlin }

\begin{table}[h]
\caption{Equipment deployed in the quantum, key-management, and application layer of the Berlin QKD-testbed architecture.\label{tab03}}
\begin{tabularx}{\textwidth}{p{3.3cm} l}
\toprule
\textbf{Layers}	& \textbf{Equipment}
\\ \midrule
Quantum layer
& \makecell[l]{
  DV-QKD systems by ID Quantique,\\
  DV-QKD systems by Toshiba,\\
  PQC key-exchange system developed by Open Quantum Safe \cite{oqs2022liboqs}
}\\ \\[-1.5ex]
Key-management layer
& \makecell[l]{
  key-management system (KMS) internally developed by DT,\\
  hardware security module (HSM) by Gemalto
}\\ \\[-1.5ex]
Application layer
& \makecell[l]{
  L1 hardware encryptors by Adva,\\
  L3 hardware encryptors by Thales
}\\ \bottomrule
\end{tabularx}
\end{table}

\begin{figure}[H]
\centering
\includegraphics[width=0.9\textwidth]{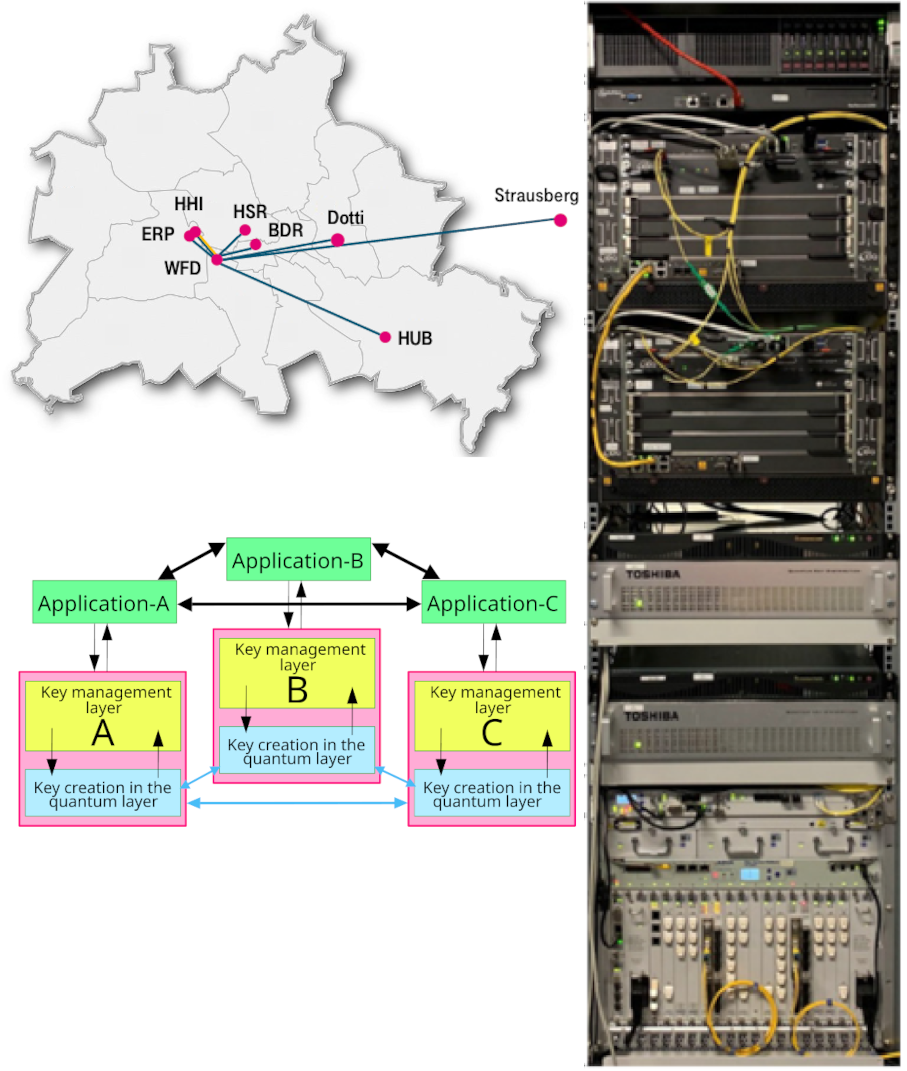}
\caption{Dark-fiber topology in the Berlin metropolitan-area testbed (top left), deployed three-layer architecture (bottom left), rack hosting QKD modules, servers, HSMs, and encryptors (right).  For more details the reader is referred to \cite{Berlin-testbed}.\label{fig01}}
\end{figure}

The QKD testbed Berlin connected several offices, institutes and network operating centers of Deutsche Telekom in the Berlin metropolitan area \cite{ICCCAS23-1} using dark optical fiber. The testbed utilized a typical QKD-network three-layer architecture consisting of a quantum, key-management, and application layer but introduced PQC key exchange in the quantum layer for hybridization purposes (as discussed in Section 1.). Table~\ref{tab03} lists the systems deployed in either architectural layer, a photo of a testbed rack hosting the equipment is shown in Figure~\ref{fig01}. 

The Berlin testbed utilized an ETSI GS QKD 014 \cite{ETSI-014} API and the encryption keys  generated by the key exchange systems were imported into a hardware security module or local key store as ``single point of trust''. Also, an ETSI 014 API exposed the encryption keys to application encryptors on the application layer. 

\subsection{QKD testbed in Poznan}
\begin{figure} [H]
\centering
\includegraphics[width=\textwidth]{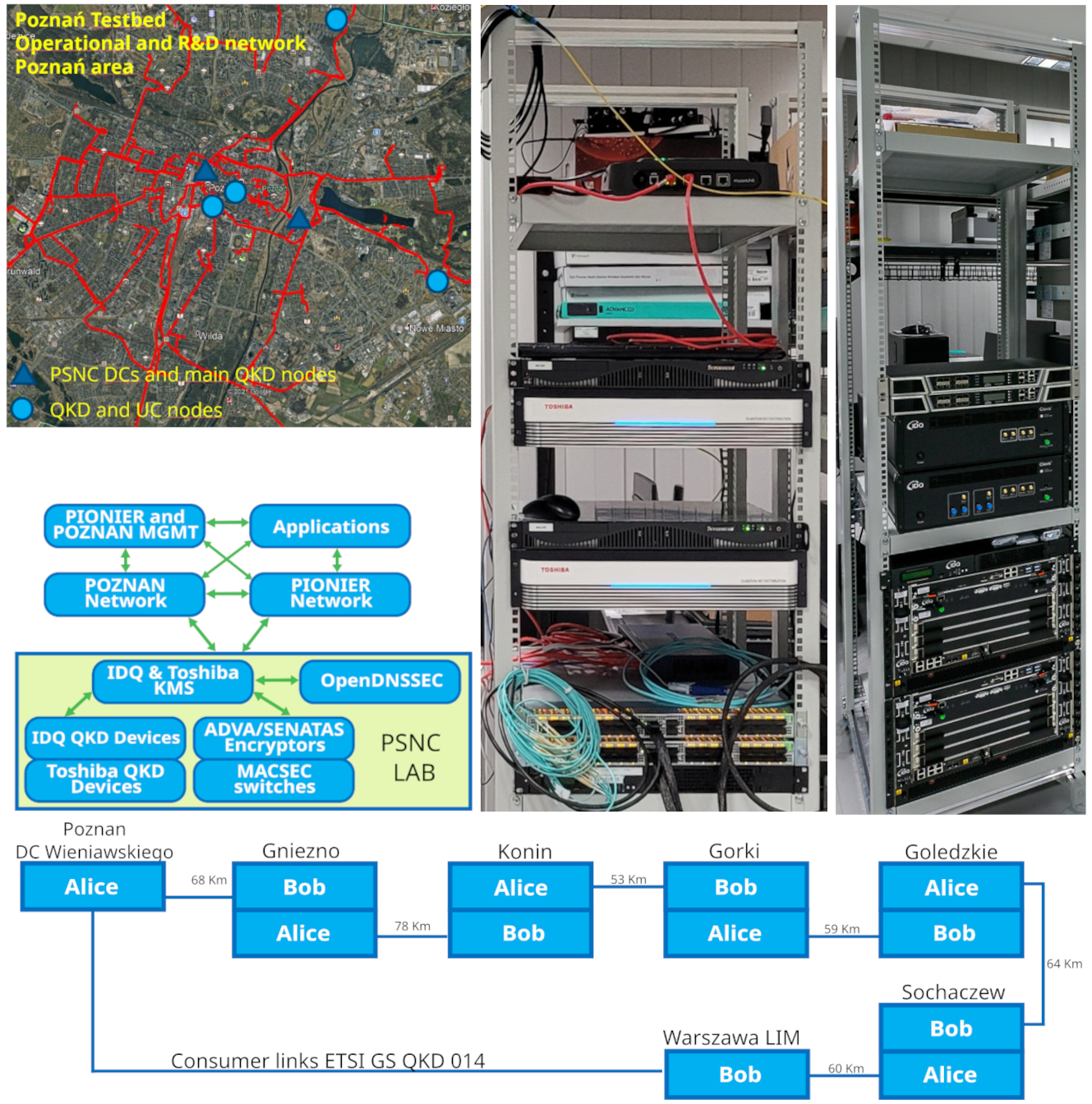}
\caption{PSNC testbed together with the dark fiber topology in the Poznań metropolitan area (top left), deployed layer architecture for both metro and backbone networks - POZMAN and PIONIER (center left), racks hosting QKD equipment and encryptors (right), additionally connected trusted-node configuration of the long-distance QKD link between Poznań and Warsaw (bottom) \cite{NLPQT}.\label{fig02}}
\end{figure}

The QKD testbed in Poznań \cite{Poznan-testbed} was implemented based on the POZMAN and PIONIER network infrastructures. Poznan QKD testbed is under integration with Polish QCI infrastructure and full publication of the infrastructure is pending. PIONIER is the Polish research and education backbone network and POZMAN is the metro area research and education network in Poznań. Both infrastructures connect multiple different research and public institutions and provide several services for its users and environment – both nationally and internationally. The PIONIER network also extends to Europe and reaches important research and education locations such as CERN in Geneva.  The QKD testbed in Poznań used mainly metro area dark fibers between PSNC offices and node locations/service hubs. For specific QKD use cases the testbed used long distance PIONIER backbone dark fibers that connected directly to PSNC lab and metro infrastructure. The testbed implemented similarly as in other locations a three layer architecture consisting of a quantum, key management and application layer, the latter being based on already existing network services. Table~\ref{tab04} lists the systems deployed in either architectural layer, a picture of a testbed rack hosting the equipment is shown in Figure~\ref{fig02}. 

\begin{table}[h]
\caption{Equipment deployed in the quantum, key-management and application layer of the Poznań QKD-testbed architecture.\label{tab04}}
\begin{tabularx}{\textwidth}{p{3.3cm} l}
\toprule
\textbf{Layers}	& \textbf{Equipment}
\\ \midrule
Quantum layer
& \makecell[l]{
  DV-QKD systems by ID Quantique,\\
  DV-QKD systems by Toshiba,\\
  PQC key-exchange system from Open Quantum Safe \cite{oqs2022liboqs}
}\\ \\[-1.5ex]
Key-management layer
& \makecell[l]{
  Key-management system by ID Quantique,\\
  key-management system by Toshiba,\\
  security module implemented with open software solutions,\\
  OpenDNSSEC
}\\ \\[-1.5ex]
Application layer
& \makecell[l]{
  L1 hardware encryptors by Adva,\\
  L3 hardware encryptors by SENETAS
}\\ \bottomrule
\end{tabularx}
\end{table}

The Poznań testbed used mainly an ETSI GS QKD 014 API and the encryption keys generated by the key exchange systems were imported into the Key Management System and subsequently directly consumed by services and/or applications or possibly by the open software security module. Also, an ETSI 014 API exposes the encryption keys to application encryptors on the application layer. As an alternative the testbed implemented also the novel ETSI GS QKD 020 API \cite{ETSI-020}, currently under development. However, due to incomplete support in hardware it was not fully functional.  The PQC protocols implemented in PSNC testbed were based on the same mechanisms than in the Berlin testbed – system and were developed using Open Quantum Safe \cite{oqs2022liboqs} This approach guaranteed best compatibility and convenience to quickly implement any changes in the configuration. The Poznan testbed was connected with a long distance physical QKD link chain Poznan - Warsaw presented on Figure 3. It is 380 km long with trusted node configuration and using dedicated dark fibers for quantum channel. At the ends of these link two consumer applications have been configured that access the generated key material using the ETSI GS QKD 014 API.
This setup allowed to connect metro and long distance (national-scale)  QKD services.  \\

\subsection{QKD testbed in Madrid}
\begin{figure}[H]
\centering
\includegraphics[width=\textwidth]{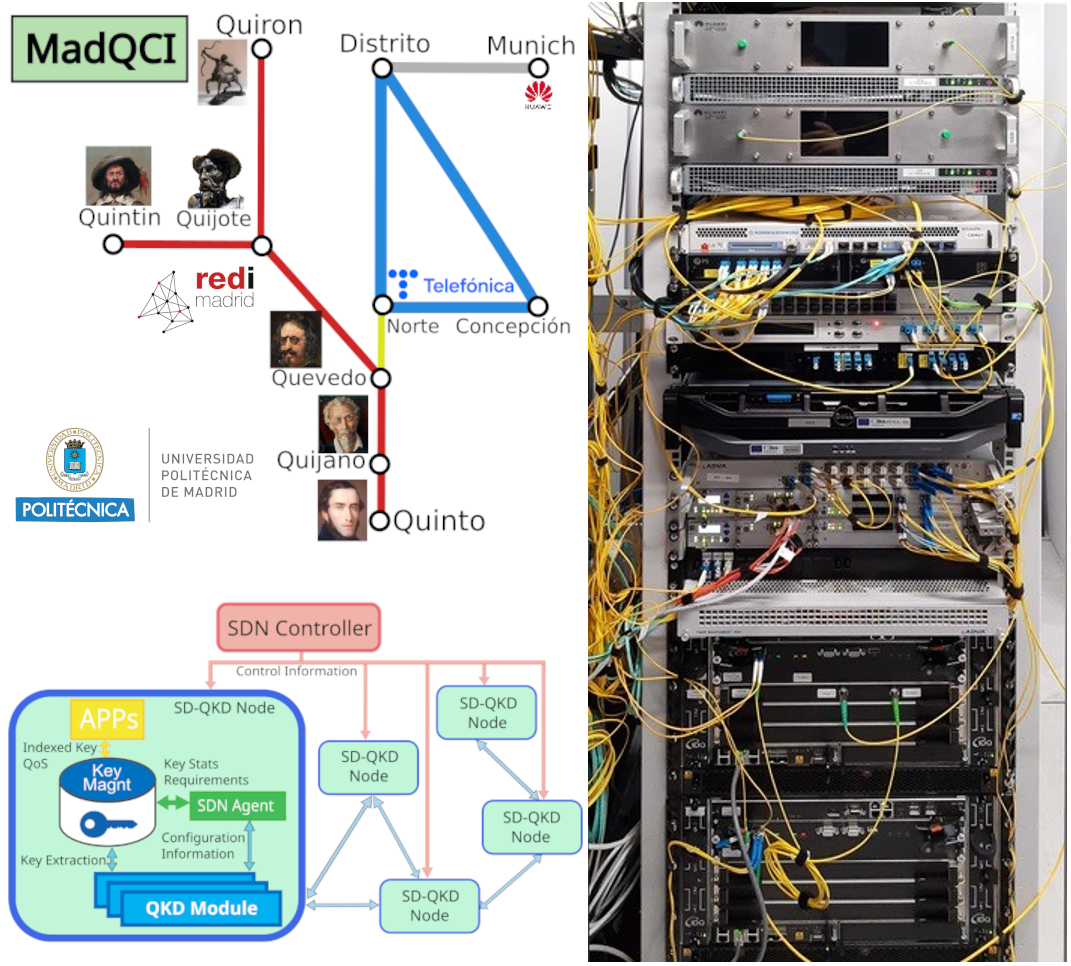}
\caption{Madrid Network – topology (top left); Madrid Network – functional diagram (bottom left); Quijote Node (right) For more details the reader is referred to \cite{martin2023madqci}.\label{fig03}}
\end{figure}

The Madrid Quantum Network or Madrid Quantum Communication Infrastructure (MadQCI) testbed \cite{martin2023madqci} has been conceived as a field trial of a real QKD production network, connecting the infrastructures of two Spanish telecoms. On one side is Telefonica, the major telecom operator in Spain, and on the other is REDIMadrid, that connects all research centers and Universities in the Madrid region. It combines dark fibers and real production channels simultaneously. We also mention that the Madrid QKD testbed logically integrates a single QKD link, situated at the Munich Research Center of HWDU, separated by  a  (direct line) distance of 1445km from Madrid. It is shown  as dot in Figure ~\ref{fig03}, connected to Madrid by a grey line.

The MadQCI follows a QKD software defined networking (SDN) design principle \cite{QKD_SDN}, allowing an easy integration of QKD required hardware such as QKD modules, encryptors, QRNGs etc. (Note that the QKD network architecture design in this case does not follow the ITU-T standards to this end - ITU-T Y.3800 \cite{ITU-T} and subsequent documents - as do the testbeds in Berlin and Poznan  but relies on the quoted SDN alternative paradigm.)   The MadQCI is based also on the Trusted Node approach \cite{peev2009secoqc} implemented in Madrid as a set of disaggregated but securely interconnected hardware and software components, each of them with clear and strict responsibilities, Figure~\ref{fig03} shows this approach and Table~\ref{tab05} lists the hardware and software systems deployed on this testbed.

As this has been done in Berlin and Poznan, the PQC infrastructure of the Madrid testbed also implements the Open Quantum Safe \cite{oqs2022liboqs} library. The Madrid PQC implementation exposes the ETSI GS QKD 004 interface \cite{ETSI-004}, with Quality of Service  based on key rate. It uses Kyber 1024 as Key Encapsulation Mechanism, but NTRU is also possible. 

\begin{table} [h]
\caption{Equipment deployed in MadQCI. Note that the layout of this table is unlike the preceding two ones, as a consequence of the differing architecture of the Madrid QKD network.\label{tab05}}
\newcolumntype{C}{>{\centering\arraybackslash}X}
\begin{tabularx}{\textwidth}{p{3.3cm} l}
\toprule
\textbf{Planes}
& \textbf{Equipment and software deployment}	\\
\midrule
\makecell[l]{Quantum\\ forwarding plane}
& \makecell[l]{
  DV-QKD systems by ID Quantique,\\
  DV-QKD systems by Toshiba,\\
  CV-QKD systems by Huawei Technologies Duesseldorf (HWDU),\\
  PQC key-exchange developed by UPM,\\
  key-forwarding and key-store software modules developed by UPM
}\\ \\[-1.5ex]
Control plane
& Modules developed for the Madrid QKD-SDN software stack by UPM
\\ \\[-1.5ex]
Application plane
& \makecell[l]{
  L1 hardware encryptors by Adva,\\
  L2 hardware encryptors by Rohde \& Schwarz,\\
  L3 software encryptors developed by UPM,\\
  further security applications developed by UPM and Telefonica
}\\
\bottomrule
\end{tabularx}
\end{table}

Note that this table differs in form from the previous one, due to the distinct QKD network architecture (a SDN one) used in MadQCI. 

For Key delivery ETSI GS QKD 004 API and ETSI GS QKD 014 API are used in a transparent way in the south and north bound interfaces of the UPM control (or intermediate) layer. It allows delivering key material to the application layer (e.g. encryptors) over appropriate interface independently of the key delivery interface used by the QKD devices.

Due to its SDN nature, the MadQCI is highly based on open interfaces and standard tools, schemes and protocols, also typically used by telecommunications industry. The same approach has been followed with respect the QKD modules, and important standards have been integrated into the network. In addition to the aforementioned ETSI GS QKD 004 and 014 APIs for key delivery, ETSI GS QKD 015 \cite{ETSI-015} is used for interfacing the  SDN-QKD node control agent and the SDN logically centralized network controller, that orchestrates any key delivery between  end-to-end (E2E) nodes. 

\section{Key-generation technologies}
In this section we address different key generation technologies and paths that we have deployed. We start with a basic description of QKD key exchange systems and their integration into telecommunication provider infrastructure. Due to multiple network and QKD providers, protocols and control systems involved in this project \cite{Berlin-testbed, martin2023madqci}, we address how to exchange keys via QKD border nodes, via quantum safe border nodes (PQC) and, for further increased security, via a quantum safe (PQC) network of disjoint network paths. Here the availability of QKD in each separate network is taken for granted. 

\subsection{QKD networks}
QKD links generate encryption keys at distant network nodes. The security of QKD links is based on the security of the QKD protocols and their implementation. The QKD systems establish (at least on a protocol level) an ITS communication channel, meaning, as already stated, that the key exchange is secure independently of the resources of the attacker\footnote{Note that sometimes ITS is defined as security against adversaries with unlimited computing resources, where the latter naturally include quantum computing ones. The most general attack by an adversary with arbitrary quantum resources can be viewed as a quantum computer plugged in the communication line (not a remote quantum computer!). In this sense the two definitions are identical although an adversary with unlimited computing resources can erroneously be understood in the more limited sense of an adversary with only unlimited REMOTE (quantum or classical) computing resources.}. Each QKD Module exposes a key delivery API, it could be a proprietary interface, or a standard one like ETSI GS QKD 004 or ETSI GS QKD 014. These interfaces allow the higher level layer(s) to store securely way all the key material generated by each QKD Module in a KMS/HSM.  In our experiments in the Madrid test bed, we combined both ETSI interfaces in a transparent manner, storing key material per node that has been delivered through both interfaces. (As already mentioned, at the Berlin and Poznan test bed only ETSI GS QKD 014 had been used.) 

As already also discussed, the QKD networks are typically based on the principle of trusted nodes (secure nodes, from which no secret information can be retrieved) \cite{peev2009secoqc}. The trusted nodes are physically connected through quantum channels (and classical associated channels). The key material generated on a such a QKD link is local to the (two) trusted nodes that each link connects, meaning these keys are only known to the legitimate users at the endpoints of this link. A QKD network is a set of (potentially dynamic) interconnected QKD links, each of them with their local key material. This means that if two end-points want to share the same key when they do not share a direct quantum link, the network needs to transport key material  utilizing the key generated over quantum links. As noted this process is typically denoted as key forwarding, key transport or key relay, and is a secure communication of the final E2E key using the key material generated per link to protect the final key transport  from node to node until the destination one is reached. Once the destination has obtained the key material, it is possible to establish secure E2E communication. This requires a precise synchronization of distributed KMS among all the intermediate nodes involved in a key transport path, for any E2E key delivery on the network. 

 Before we continue, we point out that the main (although by no means the only) application of symmetric key material is encryption, i.e. the utilisation of the key for encryption purposes.  While the One Time Pad encryption (combined with almost strongly universal$_2$ hashing -- see \cite{Carter} and references therein -- for ensuring message integrity) is ITS, this method is too "key hungry" to be readily applied. For this reason traditional block cipher algorithms such as AES or ChaCha with 256 Bit keys are widely used. While not being ITS these are currently believed to be (at least) quantum safe as the best quantum attack known against symmetric cryptography is based on the Grover algorithm for quantum computers that allows executing a brute force attack with "just" a quadratic speed-up.
 

\subsubsection{Long-distance (emulated) QKD links}
The long-distance links as of today in Europe can only be realized using a QKD emulation technology (here we use PQC methods), since, as already mentioned, neither long-distance QKD, nor trusted repeating chains and/or (constellation of) QKD trusted satellites are yet deployed on the continent. For this reason, all long-distance links have been realized using PQC KEM protocols and SIG authentication. We have, however, as discussed in the introduction chosen a two-factor approach, in which different implementations of PQC KEMs have been used and different physical paths have been employed. Specifically, different networks, including the terrestrial internet and commercial satellite systems were used for the demonstrations. The proposed approaches are shown in Figure \ref{fig06}. 

We have extended the QKD key forwarding concept to a border node problem to be able to transport key material on network segments and ``long hauls'', for example in the same metro network, but also between different metropolitan networks, even on an intercontinental scale. In this work we propose four border-node-methods: Link-based border node, Long-haul link-based border node,  Long-haul application-based border node and Long-haul application-based border nodes with multi-path diversity, that we have deployed on selected points of presence (PoPs) of the various metro networks. (Note that two metro networks - the Telefonica and REDIMadrid ones can be seen as a part of the Madrid network, which logically includes also the Munich single link. For the purposes of the present paper these three Madrid network segments can be seen as three independent networks.) 

A general remark: The four methods have a purely experimental purpose, but they may be seen as a demonstrator or rather an emulator for a trans-European long-haul QKD network. 

\paragraph{Link-based border node -- Method 1.}

This method, specifically as presented here, is appropriate for connecting QKD Networks with similar or even identical architectural design.
It has been used in MadQCI to connect the networks of two different telecommunication providers, the REDIMadrid QKD network and the Telefónica QKD network. Specifically, the link between border nodes  is the one between the nodes Quevedo (REDIMadrid) and Norte (Telefónica) and is shown in Figure~\ref{fig03} as a yellow line. It is used simultaneously by an Id Quantique system running in the 1310 nm band and a number of HWDU links running in the 1550 nm band. 

A hybridized key, produced by XOR-ing the keys generated over a pair of those direct QKD links (an Id Quantique link and one HWDU link), is shared between border nodes of the two QKD networks (MadQCI segments) and used as a bridge to transfer keys from any node in one network to any other node in the other network using the hop-by-hop approach.  (Note that this key is further XOR-ed with a PQC key between any two communicating nodes, something that appears redundant from an abstract cryptographic protocol perspective but is security relevant in view of the lower security level of implementations, as discussed in Section 1.)

From an operational perspective the link-based border node method appears to be analogous to the operation of a regular metro network QKD link. However, taking into account that the link connects two different QKD networks, we 
extended the functionality of the Quantum forwarding plane (the set of functionalities and devices required to forward the QKD keys through the network, see Table~\ref{tab05})  to be able to transport the final  keys from one administrative domain to another. This extension also includes the  control mechanisms (control planes in the QKD SDN case) of the networks. As in this case  both administrative domains are QKD-SDN based,  they have their separate respective SDN controllers, that, for security reasons, are not allowed to communicate directly one with the other, in contrast to the case of a single QKD network. The extensions stem from the need to connect the control plane and quantum forwarding plane  in one network to their counterparts in another network, and, obviously, both extensions follow the same paradigm. When an application requires secure communication with another one that is located in a different network, the source and the destination of the application are in distinct (mutually foreign) network domains. For this reason the key forwarding configuration of the nodes needs to be carried out in one network, from the source to the border node. The latter is configured as relay node to its counterpart border node in the second network. From there key forwarding is configured to the destination node.
The border nodes are the only ones allowed to communicate with external networks, and typically this operation requires a strict Service Level Agreements and a corresponding negotiation protocol. The SDN controllers of each network need to configure also all the necessary intermediate relay nodes in their respective domain, together with any resource allocation needed. Once the  key has been transported from the source node in one network, to the destination node in the other, an E2E secure communication can take place. 

\paragraph{Long-haul link-based border node -- Method 2.}


This method is logically similar to the previous one, and it is adapted to the case in which no distant QKD keys are available.
It has been used to connect two QKD networks: one is a MadQCI segment -  the Telefónica Network, and the other is the network segment represented by the remote QKD link  at the HWDU facilities in Munich. The long distance between both network segments cannot be presently bridged by true QKD links as already discussed. As an alternative and since both segments are based on the same SDN paradigm and design, a long  distance QKD link is emulated using PQC  between the corresponding border nodes at Telefonica Research in Distrito and the Munich Research Center of HWDU,
The emulated long distance QKD link runs two distinct PQC KEM protocols, the outputs of which are hybridized as in Method 1. to create a border node - to border node key (i.e. through the emulated distant QKD link).
The rest of the process is strictly analogous to that described in Method 1.
We point out, however, that a modification of the SDN control mechanism (the control plane) is needed. Specifically, ETSI GS QKD 015 \cite{ETSI-015} needs to be overridden and extended by a “link type” property that can be QKD, PQC, RAW (i.e.traditional communication) or other.
Essentially, the SDN node control instance (the SDN Agent \cite{ETSI-015}) manages this as another QKD link, but it knows that this is a PQC link. The ETSI GS QKD 015 has been extended to give support for this specific feature, adding the property “link type”, that it can be QKD, RAW, PQC or other. 
 This is the most general description of Method 2, which is similar to Method 1.

In the present realization, however, for simplicity, and to demonstrate that different domains could be easily merged through the SDN approach, we have decided to manage the Telefónica and  Munich segments from a unified SDN  point of view. This means that there is only one SDN controller managing all the QKD Trusted Nodes in the Telefónica segment and the Munich link as if this would be the same physical network. Such approach  simplifies the final E2E key delivering and demonstrates the potential operation in case of availability of meta- (or cross-border) controllers.

We further decided to implement this link using  the ETSI GS QKD 004 key delivery interface  because of the quality of service options offered by this API. It allows us to emulate this link choosing a constant key rate. We selected 256bps and a key length of 32 bytes. We point out that even this conservative speed choice allows us to simulate a set of trusted nodes with key provisioning. The PQC algorithm used for this type of link is presented in detail in the discussion on the subsequent Model. For implementation, again, the  Open Quantum Safe \cite{oqs2022liboqs} library has been selected.

\paragraph{Long-haul application-based border node -- Method 3.}
This approach is oriented to the interconnection of networks that are different in terms of their design, for example: networks based on the  traditional QKD Network layered architectere (Berlin, Poznan) with a network based on a different architectural  paradigm such as, e.g., QKD software defined network (Madrid). (This approach has also been used to connect Berlin and Poznan, although Method 2. would also have been possible.). The idea is to have an application service running on authorized nodes of each network. This application service is administrated in each domain by the respective network operator. Due to the long distance between the testbeds, both sides of the application establish an emulated QKD link (a PQC link instead of a QKD one), albeit in a two-factor manner, which is based on hybridization of two implementations but over the same path. In detail, the algorithm consists of the following steps: 

\begin{enumerate}
\item	Retrieve random numbers (RNDs) and matching identifiers RNDIDs from a (quantum) random number generator ((Q)RNG). 
\item	Encrypt the random numbers using different PQC KEM  algorithms, independently and hybridize the outputs to a single key string.  
\item	Add meta information, for example the RNDID, a key validity period or the name of the sending node is added to the encrypted random number. The entire package is finally signed using a different PQC SIG algorithm for each KEM choice.  
\item	After data serialization, the data package is sent from one border node to the other border node.
\item	Finally, the sending side pushes the encryption key, the identifier and the corresponding meta data into a KMS that may be a HSM or an encrypted file share.
\end{enumerate}

At the receiving node, the key exchange protocol consists of the following steps 

\begin{enumerate}
\item	After the reception of a message the sender is identified by its IP address. 
\item	The message is deserialized and the signature of the sender is validated using the PQC SIG algorithms. The appropriate public SIG key(s) of the sender is (are) determined from the IP address recorded at step 1.
\item	If validated, the receiver de-hybridizes and subsequently decrypts the encrypted random number, the identifier and corresponding metadata.
\item	Finally, the receiving side pushes the transported key, the identifier and the corresponding meta data into an KMS.
\end{enumerate}

Since the algorithm is applied on the application level, no low-level details of the network architecture are required for this service. The border node service is running constantly, and it stabilizes a constant PQC link where randomized keys could be obtained by several methods (one of them being the use of a QRNG).  The keys are protected and shared between both sides of the border-node to border-node link and used as transport keys to protect the final E2E key material. Note that if (one of) the SIG or KEM implementations is broken this might result in a denial-of-service attack as the final keys between the distant nodes might not coincide or the messages be considered unauthorized. 

 A quantum-safe long haul application-based border node has been implemented as a regular QKD key consumer application on top of the KMS layer  using ETSI GS QKD 004 (ETSI GS QKD 014 could also be appropriate). This approach has been designed for the key exchange between different (architectural and functional) networks. Applying the key interchange on the application layer hides the low level details of the network and delegates the key transport to the operators network level. Indeed, the mentioned connectivity application only needs to receive the secure key material to transport, independently on how this key material has been created or transported in each network. Note that with this approach, the unique requirement for the QKD network is the availability of a key delivery interface on each QKD border node, to obtain the  key before a quantum-safe KEM is used to encrypt the messages sent  to the application in the adjacent network.  Note that no conflict occurs if the interfaces in both networks are not the same.


For example, assume that one network is a realization of a QKD SDN architecture and the second follows a layered architecture QKD network  design. In the SDN QKD network, the request to transport  key  between the QKD Nodes on both sides of a link/chain (source and destination), is directed to the network-internal SDN Controller. The SDN Controller organizes all the intermediate QKD Trusted Nodes in relay mode to transport the key up to the border node, providing all necessary resources. As a result, the secure key is ready on the application's side of the first network border node. Then it is transported to the border node of the second network, decrypted and injected into the KMS of the layered network, which  takes over and negotiates E2E keys to  transport the keys to the final destination. (This is in contrast to  controller assisted transport in a QKD SDN network.) At this point E2E secure communication from one network to the other can be carried out. Note that this application could also be implemented as a separate application that receives keys (for example in one port) irrespective of their origin.

In future, when direct QKD links between the border nodes of different QKD Networks will be available, (one of) the consumer-application(s) supporting this Method, will simply utilize the direct QKD link key as key source and encrypt the payload-key by a symmetric ITS method - i.e One Time Pad.

\begin{figure} [H]
\centering
\includegraphics[width=0.6\textwidth]{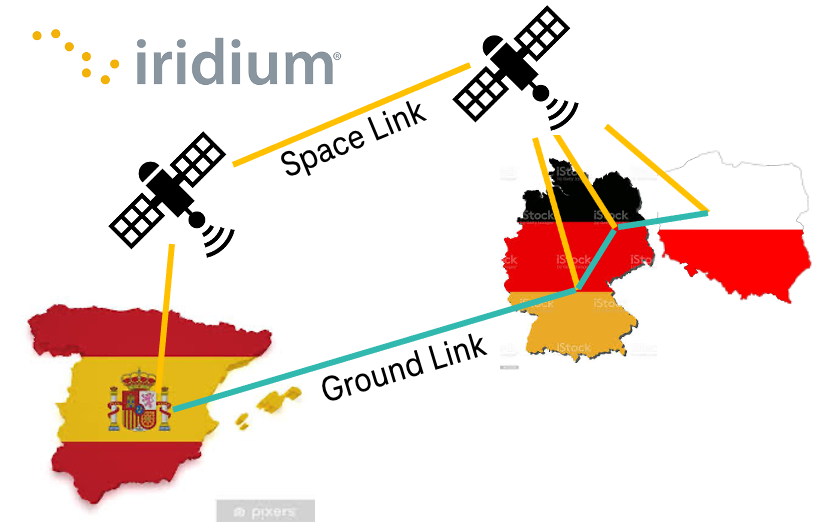}
\caption{The disjoint network is realized by a ``space link'' via the Iridium network and a ``ground link'' via the public internet. The network connects the gateway nodes of the Madrid, Berlin, and Poznan QKD testbeds.  The Munich Research Center of Huawei in Germnay serves as a pseudo-internal node of the Madrid network.\label{figSatellite}}
\end{figure}

\paragraph{Long-haul application-based border nodes with multi-path diversity -- Method 4.}
This approach adds multi-path security to the previous one using two disjoint network links. We have used the public internet or the ``ground link'', and a satellite-based link via the commercial Iridium network, or the ``space link''. We chose the Iridium network, because it is a commercial network with worldwide coverage and affordable access. The setup is shown in Figure~\ref{figSatellite}.  Note that in the introduction, we have already mentioned that the two-path diversity is preferable security-wise. Again, the long distance forces us to use emulations of QKD. A two-path algorithm, based on different PQC KEM and SIG protocols is detailed below. It differs from the one outlined in the previous approach in the sense that non-coinciding random number strings are sent in this two-factor version along different routes and subsequently these are combined, rather than encrypting a single random string with different methods and then hybridizing the result. Similarly,  if (one of) the SIG or KEM implementations is broken this might result in a denial-of-service attack as the final keys between the distant nodes might not coincide or the messages be considered unauthorized. 


The PQC-based, two-path key exchange protocol involves the following steps for the sending node: 

\begin{enumerate}
\item	Retrieve two random numbers (denoted by $\text{RND}_1$ and $\text{RND}_2$) and a matching identifier per RND (denoted by a single RNDID) using a (Q)RNG. The index 1 is liaised to the space link, whereas the index 2 denotes the ``ground link''.  
\item	Encrypt the random numbers using different PQC KEM (Key Encapsulation Mechanism) algorithms. Depending on the chosen path, a different public KEM key is applied. Any appropriate KEM algorithm may be used. We used Kyber on the space link and NTRU on the ground link, because these algorithms proved good performance with current implementations \cite{PQC-Performance}. 
\item	Meta information, for example the RNDID, a key validity period or the name of the sending node is added to the encrypted random number. The entire package is finally signed using a different PQC SIG algorithm for each network path. Any appropriate SIG algorithm may be used. We used Falcon on the space link and Dilithium on the ground link.
\item	After data serialization, one key package is sent via the space link, the other key package via the ground link. 
\item	On successful data transmission, the sending side combines the two random numbers $\text{RND}_1$ and $\text{RND}_2$ to compute an encryption key (KEY) using a key derivation function (KDF), so that KEY = KDF($\text{RND}_1$, $\text{RND}_2$, PSK), where PSK is some (possibly empty) pre-shared key string. Any KDF standardized by NIST or ETSI may be chosen \cite{kdf, ETSI-TC-CYBER}. (In case real QKD links, and not only emulated QKD links are used, this choice of KDF  must be restricted to functions that ensure the epsilon-composability of the output; e.g., a simple XOR-ing function can be considered.) The KEYID is set identical to the RNDID and will be required for key negotiation protocols. Finally, the sending side pushes the encryption key, the identifier and the corresponding meta data into a KMS.
\end{enumerate}

At the receiving node, the key exchange protocol consists of the following steps 

\begin{enumerate}
\item	After the reception of a message on either network $\text{path}_i$, i={1,2}, the sender is identified by its IP address.
\item	The message is deserialized and the signature of the sender is validated using the PQC SIG algorithm. The appropriate public SIG key(s) of the sender is (are) determined from the IP address recorded at step 1.  
\item	If validated, the receiver decrypts the encrypted random number $\text{RND}_i$, the identifier $\text{RNDID}_i$ and corresponding metadata using the PQC algorithm $\text{KEM}_i$.  
\item	The decrypted random numbers, their identifiers and metadata are then sent to a queue for further processing.
\end{enumerate}

If two random numbers $\text{RND}_1$ and $\text{RND}_2$ with the same identifier RNDID are found in the queue, the numbers $\text{RND}_i$ will be combined so that KEY = KDF($\text{RND}_1$, $\text{RND}_2$, PSK) is computed (see above the comments on the KDF  choice). Finally, the receiving side pushes the final key, the identifier and the corresponding meta data into a KMS.  This solution makes use of Open Quantum Safe (OQS) \cite{oqs2022liboqs} for the implementation of the PQC KEM and SIG algorithms. The PQC algorithms are compiled into openssl. A PQC enabled version of the nginx web server with multiple workers is used and the python code uses multithreading to increase the performance.  

The security system preserves the secrecy of the final key, as long as a single path of the disjoint network paths remains secure. This means that even if a PQC algorithm used on one of the paths will be successfully attacked in the future, the other PQC algorithm acting on the disjoint path (if not also broken) would guarantee the overall security of the system (up to a denial-of-service, as already discussed). The KDF is combining the two random numbers in such a way, that even the knowledge of one of the two RNDs would not allow the adversary to compute the resulting final key \cite{Robust-Combination}, a property known as robust combination. As discussed, the security of the solution can even be increased by adding more disjoint paths and securing the key exchange using different PQC algorithms. Besides the combination of satellite and terrestrial networks, there are other commercial networks which are disjoint: for example, the networks of competing mobile network service providers or European research fiber networks or commercial fiber networks. Once existing, even a satellite QKD or long-haul quantum optical key exchange link may be added \cite{ICCCAS23-2,ICCCAS23-3} to make the solution information theoretical secure (albeit by a more careful selection of the KDF, as mentioned) with a significant side channel reduction. 

The long-haul application-based border nodes have been implemented in the Berlin, Poznan and Madrid testbeds on specific QKD Trusted Nodes, using physical servers. The Iridium "space network" and the "internet" are used to exchange the encryption keys over disjoint network paths. Using a virtual machine with 2 CPUs and 16GB RAM, the presented software solution manages to transfer 4kBits (which corresponds to 16 AES-256 keys per second). The virtual machine was equipped with PQC-enhanced web server and client applications running the key exchange. To do so, and to compensate for the long latency of about 600ms per request, the best performance was achieved when sending blocks of 50 to 75 keys per https session. The bottleneck of the implementation turned out to be the recombination function to grab the two random numbers and apply the KDF. We believe that the software performance can be tremendously improved by choosing a more powerful coding language, like RUST or C and by changing the software architecture to asynchronous queuing with multiple stateless microservices acting. The solution itself is scalable with more microservices exchanging keys between the end point sharing the capacity of a larger number of satellite access antennas. The usage of a LEO (Low Earth Orbit) satellite constellation, like Star Link \cite{Starlink} may also increase the performance of the implementation. Geographically, the solution may be scaled globally by integrating more end points and using existing or coming satellite constellations in conjunction with ``standard'' internet connections. 

\subsubsection{Experimental results}
The following section summarizes the Key Performance Metrics, 
of the key distribution channels between the border nodes of the three testbeds.
Each border node (as in fact any Trusted Node) was equipped with a computer system and software to run the PQC-enabled key exchange protocols. The networks have been linked by a logical, classical VPN network.

\begin{table} [h]
\caption{Result summary of the3 four key exchange methods to integrate the QKD testbeds. \label{tab:results}}
  \centering
  \begin{tabular}[t]{p{1cm} p{3cm} p{4.5cm} p{2.3cm}}
    \hline
    \textbf{Method} & \textbf{From-To} & \textbf{QKD/PQC Algorithm} & \textbf{Key Rate} \\
    \hline\hline
    1 & REDIMadrid to Telefonica & QKD \& Kyber/Falcon and NTRU/Dilithium & QoS based on 256 Bit/s \\ \hline
    2 & Madrid to Munich & Kyber/Falcon and NTRU/Dilithium & QoS based on 256 Bit/s\\ \hline
    3 & Madrid to Berlin, Madrid to Poznan & Kyber/Falcon and NTRU/Dilithium & QoS based on 256 Bit/s\\ \hline
    4 & Berlin to Madrid, Berlin to Poznan & Kyber/Falcon and NTRU/Dilithium & 4kBit/s \\
    \hline
  \end{tabular}
\end{table}

Table~\ref{tab:results} shows the four key distribution methods, their deployment location and the name of the key encapsulation and signature algorithms applied. The key exchange rate, 
is shown. 
It is important to notice that the (classical) PQC key exchange protocol makes use of TCP/IP, which corrects transmission errors by design. 
It is also important to mention that the key exchange metrics originate from a single working application or thread. By adding more CPU threads to the key exchange application, the performance can easily be scaled. Therefore using a  CPU with a higher number of threads yields higher key exchange rates.

\begin{figure} [H]
\centering
\includegraphics[width=\textwidth]{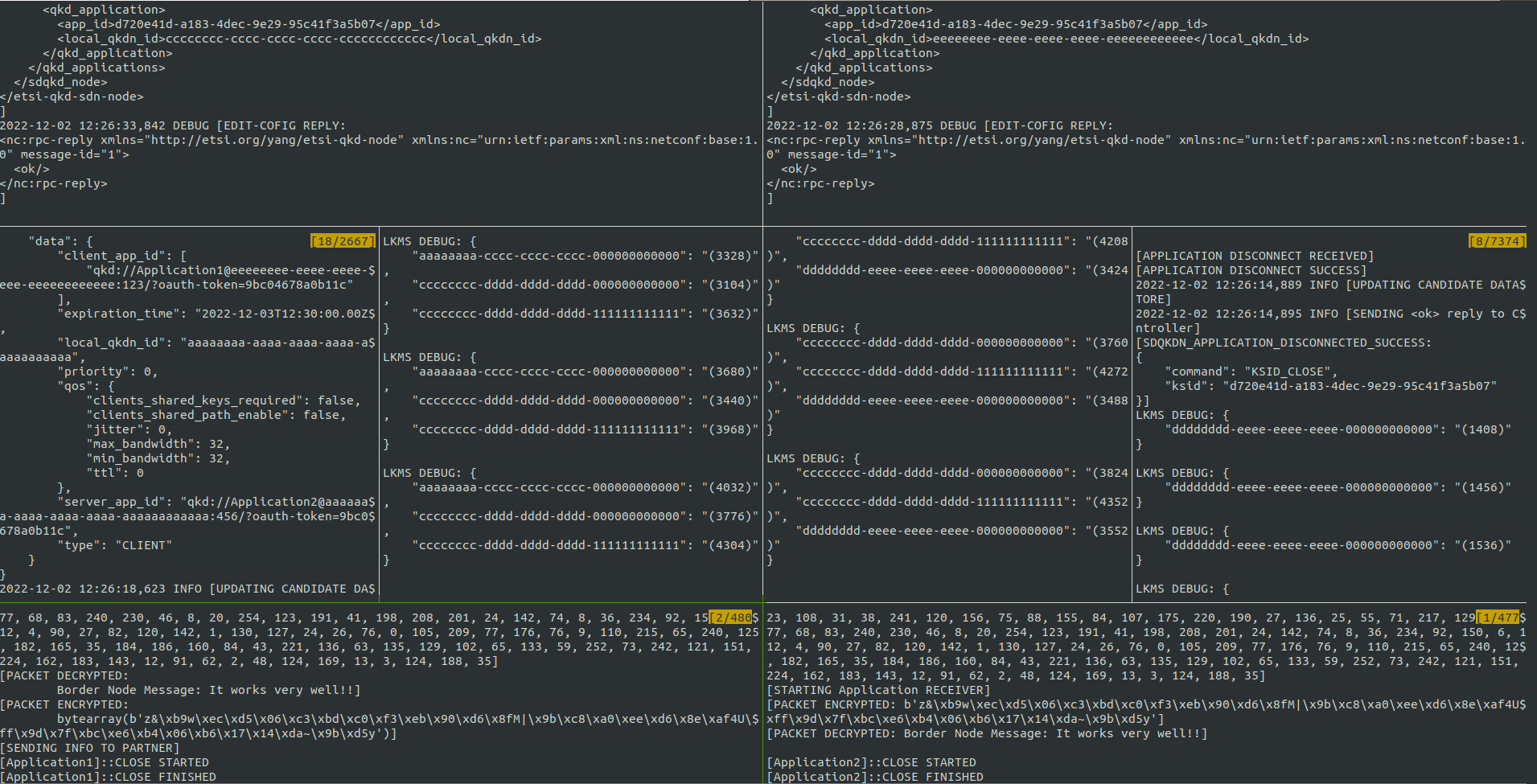}     
\caption{Application key transport between REDIMadrid domain and Telefónica domain through a QKD Link-based border node. The left part represents the REDIMadrid domain, the right part represents the Telefónica domain. The upper part are the SDN controllers of each network, both controllers are NETCONF based. The central part represents the LKMS of each node involved in the communication, on the left the source node of REDIMadrid receiving a QKD key request through ETSI GS QKD 004, in the center-left there is the border node in the REDIMadrid side, on the center-right, the border node of the Telefónica side and on the right, the destination node of the communication, showing the Application Disconnect. In the lower part, the left side represents the source application and the right side the destination  application. 
\label{fig:results-method1}}
\end{figure}

\paragraph{Demonstration of method 1}
The following figure shows a successful E2E key exchange between nodes in the REDIMadrid segment and Telefónica segment of the Madrid testbed. The left side of Figure~\ref{fig:results-method1} shows the main entities involved in this communication on the REDIMadrid side. The right side shows the Telefónica side. On top  the respective SDN controller are to be seen. These govern each respective domain, whereby each of them  does not have details of the other network. In our experiment, there are two nodes in each network that are involved in the communication, an internal regular trusted node and the border node: Sansa and Rickard respectively on the REDIMadrid side, and Lyanna and Eddard on the Telefónica side. Their respective LKMSes and associated QKD links are shown in the middle of the figure. An application asks for key in each respective network, and each SDN controller orchestrates all the nodes, including the border node, to ensure a successful E2E key delivery, as  shown in the lower part of the figure.

\paragraph{Demonstration of method 2}

\begin{figure} [H]
\centering
\includegraphics[width=\textwidth]{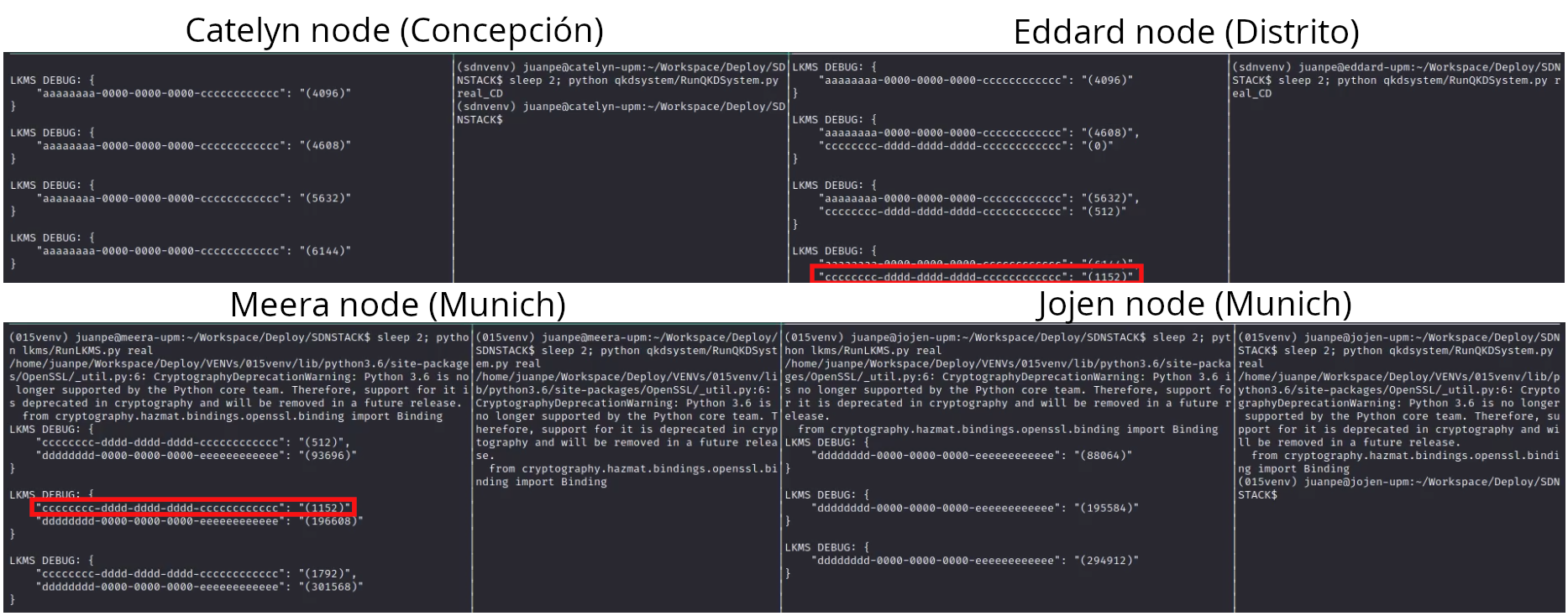}
\caption{Links between Concepción, Distrito and Munich. The QKD simulated link using PQC between Distrito and Munich is indicated in red. 
\label{fig:results-method2}}
\end{figure}

On Figure~\ref{fig:results-method2} the links and the amount of key material stored per link between Meera and Jojen nodes on Munich (link: dddddddd-0000-0000-0000-eeeeeeeeeeee) and between the Catelyn node in Concepción and Eddard node in Distrito (link: aaaaaaaa-0000-0000-0000-cccccccccccc) are shown. The long-haul link-based border node connection is operating on the link cccccccc-dddd-dddd-dddd-cccccccccccc. In the figure this link is indicated in red. It connects Eddard in Distrito with Meera in Munich with a QKD simulated link using PQC. 

\paragraph{Demonstration of method 3}

\begin{figure} [H]
\centering
\includegraphics[width=\textwidth]{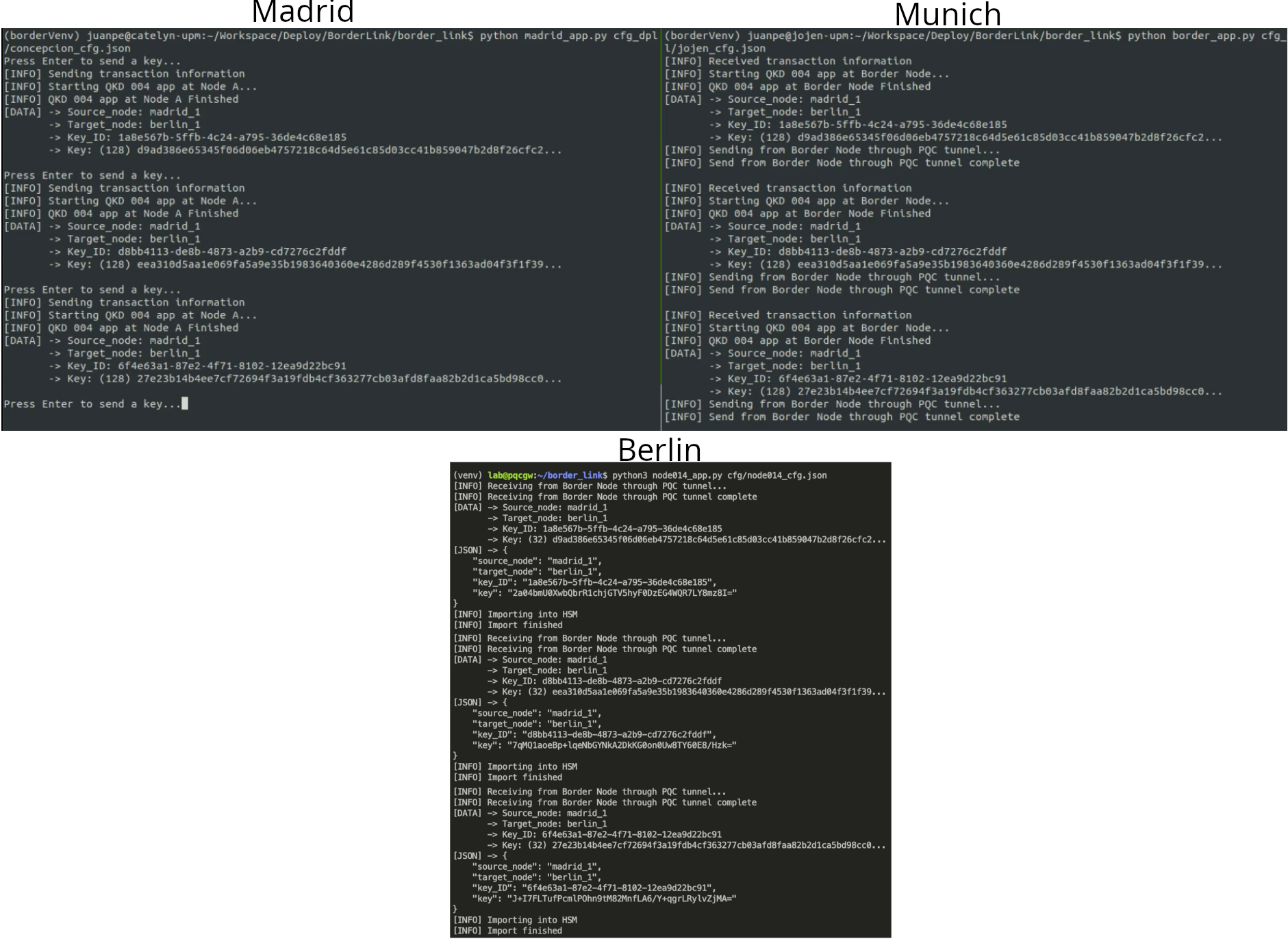}
\caption{On the left side of the image an application is started to send key material from Madrid to Berlin. 3 different keys are sent from Cathelyn node in Concepción to Jojen node in Munich where the application-based border node is running and sending the key material to the Berlin border node using a PQC link. On the lower part of the image, the key material received in Berlin is shown.
\label{fig:results-method3}}
\end{figure}

Figure~\ref{fig:results-method3} shows screen views of the full key transport from the Cathelyn node in Concepción to the Jojen node in Munich and the final transmission from there to Berlin. On the left side of the figure, a QKD application is started to communicate form Madrid to Berlin. The key transport includes a first QKD link from Concepción to Distrito, then a Long-haul link-based border node (Method 2) from Distrito to the Meera node in Munich. The third link is again a QKD link in Munich from Meera to Jojen, where the application-based border node method is realised: the received key material is encrypted by two different key encapsulation algorithms and sent to Berlin. On the Berlin side, the reception side of the application is waiting to receive the key material and distribute it inside the network. 

\begin{figure} [H]
\centering
\includegraphics[width=\textwidth]{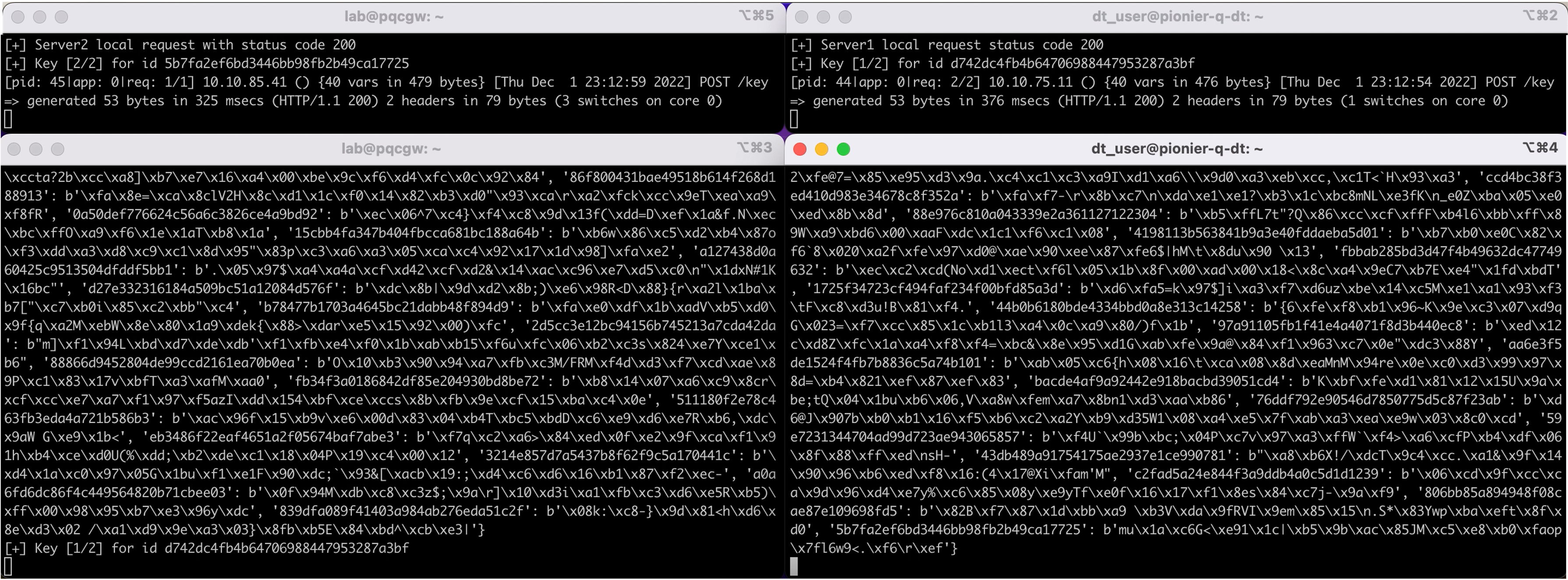}
\caption{Bi-directional exchange of encryption keys between the Berlin and Poznan border nodes. The Poznan server (top, right) and the Berlin client (bottom, left) are exchanging keys. The server log states the reception of the ground link transmission (Keys 1/2), so does the client log. In this state of the exchange process, the space link keys transmission has not yet been finalized.
The Berlin server (top, left) and the Poznan client (bottom, right) are exchanging keys (in the opposite direction). The server log already states the full reception of the keys of the second (space) segment, while the client shows the final (KDF-combined) block of 50 keys and their key identifier respectively.
\label{fig:results-method4}}
\end{figure}

\paragraph{Demonstration of method 4}
Figure~\ref{fig:results-method4} shows a screen view of the method 4 key exchange between the border nodes of the Berlin and the Poznan QKD network. On each border node, a server and a client application is launched. The server-side application is listening for exchange requests of random numbers (RNDs). The client-side application transmits blocks of 50 PQC-encrypted and signed random numbers and identifiers via the ground link (called Keys 1/2 in Figure~\ref{fig:results-method4}). At the same time, the client-side application sends an equally sized, but different block of PQC-encrypted and signed random numbers and identifiers via the space link (called Keys 2/2 in Figure~\ref{fig:results-method4}). 

\noindent
The server is required to store the keys exchanged on the ground link for a certain time period to manage for the longer latency of the satellite channel. This is realized via queues in the memory of the server. A microservice finally queries and combines the related keys using a key derivation function. The client application needs to manage the speed of the key transmission process in accordance with the network capacity. As a result, the same set of random bits is exported to the local key store of the border nodes by the client and server-side applications. 

\subsubsection{Key forwarding through border nodes}
\begin{figure} [H]
\centering
\includegraphics[width=\textwidth]{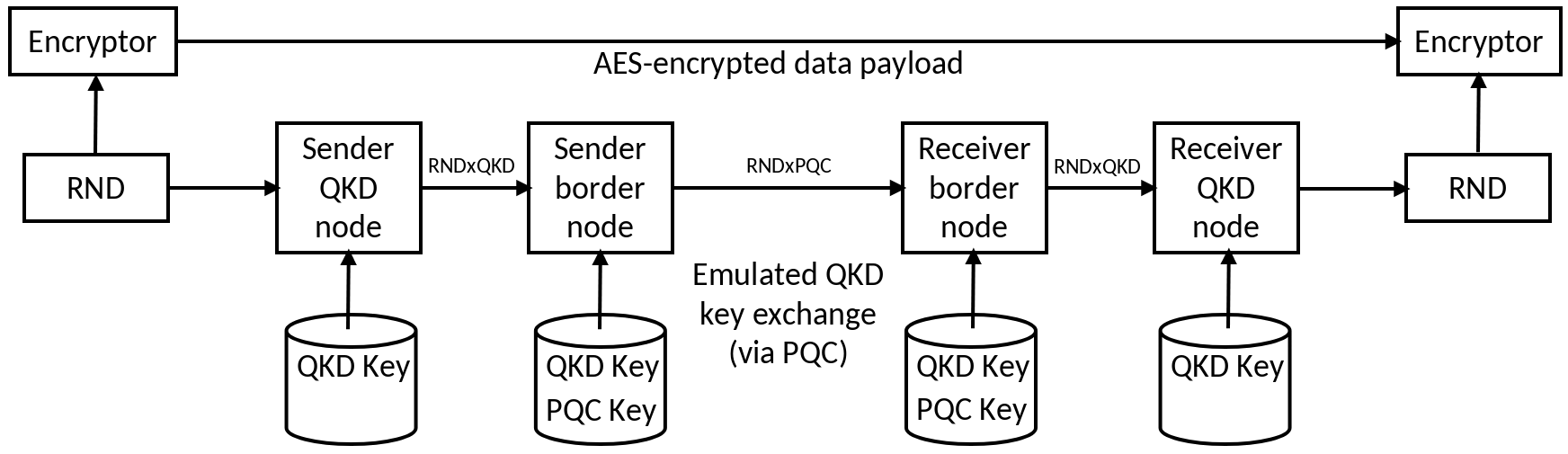}
\caption{Integration of an emulated QKD key-exchange system into a QKD architecture.  The keys of the emulated QKD key exchange are stored in the local key store of the border nodes.  A sending QKD node forwards a random number through the trusted-node chain of the senders border node and the recipients border node to the receiving QKD node.  The random number is either directly used as a final secure key 
or 
two random numbers, transferred across disjoint network links, are combined using a KDF for final secure key.\label{fig05}}
\end{figure}

This paper presents various solutions to realize a border node key exchange between individual European QKD deployments making use of QKD  and PQC as the mentioned emulation of QKD. This yields an overarching architecture, where border nodes are deployed to interconnect QKD deployments. As shown in Figure~\ref{fig05}, random numbers are forwarded from a source QKD node to the border node (communication secured by QKD keys), from there to another border node of a target QKD infrastructure (secured for now by an emulated QKD, or a QKD satellite in the future) and further on to a target QKD node. As a result, the source and the target QKD nodes share the same random number, which they may utilize as, e.g., a secure key to, e.g., encrypt the classical data payload. As described in the introductory section of this paper, the distant link (border node integration) was realized using PQC links due to the lack of appropriate quantum technology.  Clearly, the emulated QKD links  stand for true QKD ones in the context of a final architecture blueprint. 

\begin{figure}[H]
\centering
\includegraphics[width=\textwidth]{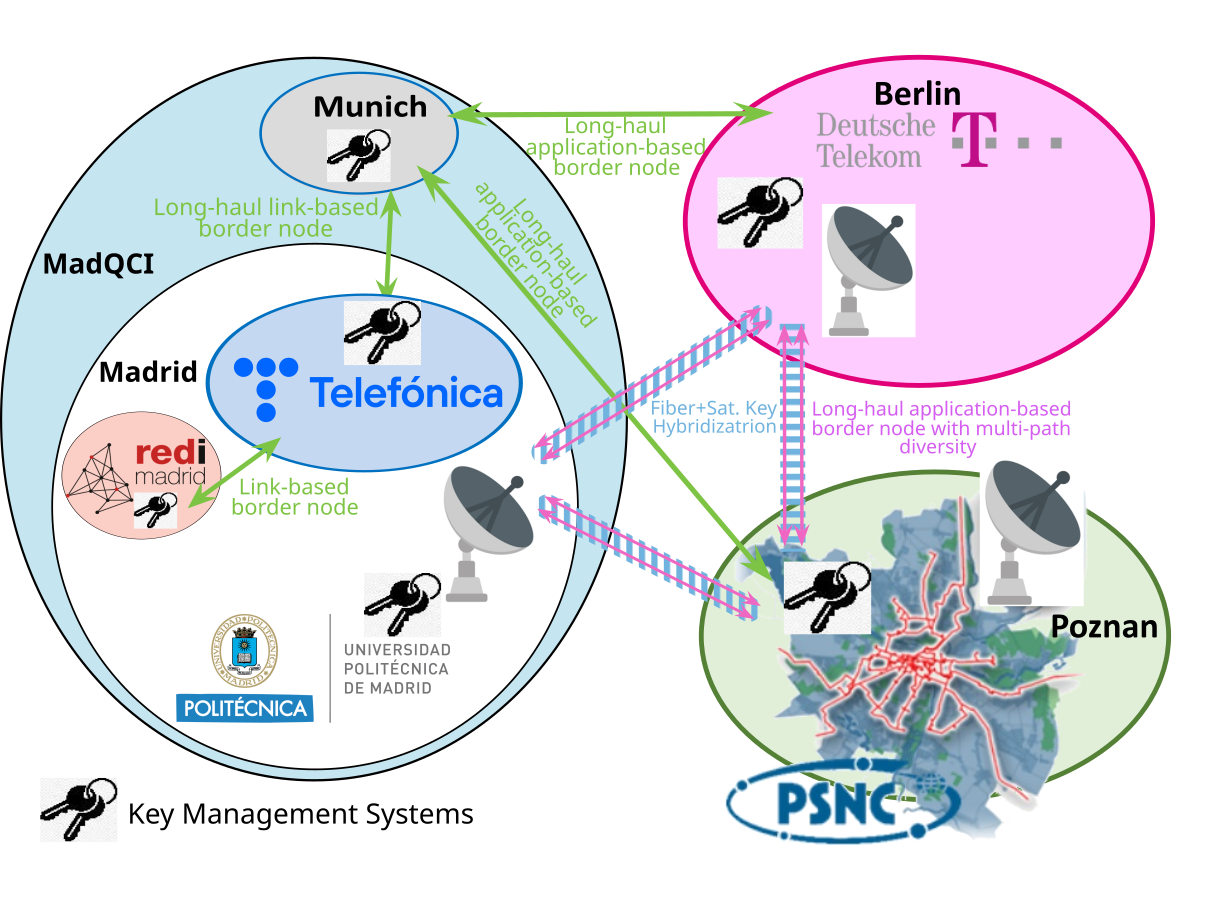}
\caption{Long-distance (emulated) QKD links connecting the metropolitan-area QKD networks of Madrid \& Munich, Berlin, and Poznan.  
\label{fig06}}
\end{figure}

The resulting network is fully meshed. Each gateway node operates a PQC key exchange server that is listening for incoming connections. Additionally, each gateway node could initiate a key exchange through the key exchange client, yielding a bi-lateral key exchange. The integration of the PQC key exchange works through standard interfaces like any other key supplier. The testbeds integrate the PQC key exchange by pushing the keys, their identifiers and their metadata directly into the KMS at every location, where they can be consumed by encryptors or applications via the ETSI GS QKD 004 or 014 API \cite{ETSI-004,ETSI-014} and an appropriate key negotiation process. Figure~\ref{fig06} shows an overview of the four border node solutions deployed in this project. (Note that the application based long-haul border node connection is always realized following the multi-path diversity approach.)

Here a few general comments on the overall standing of all four proposed methods. We underline that these were adapted to the present situation, when long distance QKD links do not yet exist in Europe, whereby these could be applied also if the availability of long distance QKD is scarce. The situation changes if  long distance QKD is available. We would first point out that in this case Method 1 will be identical to Method 2. 
On the other hand, generally, Method 1 and 2 can also be applied to dissimilar networks if appropriate extension of the work, presently performed on ETSI GS QKD 020, is carried out. Possibly, however, this is not needed, as one can always apply Method 3. In this sense the prospective role of Method 3, depends on decisions to be taken in standardization bodies. Method 4 details the multi-path diversity extension that (as mentioned already in Section 1.) can always be used to enhance security.

As far as table~\ref{tab01} in Section 1 is concerned, we draw the attention of the reader to the fact that we  recommend the utilization of row number 3 there. Until  long distance QKD will not be available to a sufficient extent we will be forced to the situation represented by row number 1 in table~\ref{tab02}. For the same reason, for the long distance connections we are using in this work we rely on the approach of table~\ref{tab01} row 2. Note independently that if Method 1 would only rely on QKD we would be using row 1 of table~\ref{tab01}.

\subsection{Hybridization of all generated keys}
Security usually is a crucial parameter in network communications. We therefore propose to integrate a hybridization scheme to compute hybrid final keys at each network node. Previous approaches \cite{aguado2017hybrid} combine PKC keys with QKD and indicate the possibility to use even PQC. In the present work the hybridization is i) following the recent Muckle scheme  \cite{Muckle} that puts forward a hybrid method for authenticated key exchange (specifically it is based on a secure hybridization, i.e.combination, of several keys of QKD, PQC and PKC origin, and the additionally, authentication of the key distribution using PSK, instead of  PQC SIG), and, ii) additionally, is relying on extended hybridization involving multiple media/paths. We are also explicitly motivating this strategy by a careful analysis of the potential security benefits of combining practical implementations of protocols (not to be confused with the mathematically formulated protocols themselves).

The key exchange module between PoPs has been modified to manage not only QKD keys, but also PQC ones. The systems also allow the use of several key exchange modules in parallel, even with the same PoPs. The KMS systems receive the keys (QKD and PQC or any other key exchange mechanism) in a transparent manner from these key exchange modules, so the KMS can establish different quantum safe key exchange session with other PoPs, using QKD links, PQC links and a combination of both in a simple way. 

To do that, the key exchange modules are running in parallel on each node delivering keys to the KMS. The payload functionality of the key exchange modules is vendor- and technology-independent, and the only exposed interfaces are the ETSI GS QKD 004 or 014 ones. The KMSes only get notice of a new link between two PoPs. This design enables generic integration of additional links,  and the keys generated by PQC are internally managed by a KMS exactly as any key material generated by QKD. All these interfaces can be seen as quantum safe key delivery interfaces, whereby (Q)RNG serves as key source \cite{herrero2017quantum}. The PQC link is implemented as a TCP connection, with package delivery granted and in order. That allows to simulate a potentially infinite, protected stream of keys with PQC between any two peers of the network and identity authentication of the end points is required only when a new TCP stream is started. The use of PQC links allows to have long distance quantum safe links where QKD cannot reach right now, but the current design, exposing standard QKD interfaces, makes possible to replace PQC by QKD links when the technology will be mature enough and with a minimal impact on the rest of the architecture. The hybridization of QKD and PQC keys will, as described earlier, deliver a key exchange system with significantly reduced side channels due to the use of principally different technologies and implementations thereof. 

The key hybridization process is done applying a hybridization KDF. The key hybridization process in performed as an internal process on the KMS that manages different internal key stores. The key hybridization process needs to process the appropriate key bits so that a new, hybrid key is computed, stored and handed over to applications and encryptors. Alternatively, hybridization may be left to the application, since the application oversees enforcing the required security level itself – by picking a key or a combination of keys exchanged under the right security paradigm.

\section{Conclusions}
QKD networks are a key building block for quantum-safe communications and the interest in related research fields and subsequent industrialization is increasing rapidly.  There is an emerging necessity of deploying QKD metro networks, but also connecting these metro networks over long distances.  The present publication demonstrates several, viable approaches for a quantum-safe key exchange between three of the major, production-grade QKD testbeds in Europe, that are Berlin, Madrid, and Poznan.  These three testbeds are different in terms of network architecture, functionalities, and management, reflecting the different ways that a telecommunications company may operate its infrastructure.  Different key exchange realizations were defined, including key exchange over different physical media (fiber and satellite), to enable E2E communication between all nodes of all participating networks.

A cross-European, E2E key exchange was designed, which follows the SDN principles adapted to QKD.  This approach allows to interconnect nodes not only belonging to the same, but also to different networks by QKD-powered, quantum-safe links.  As an example, two network domains in Madrid, REDIMadrid and Telefónica, were connected through an SDN-based QKD layer.  This approach was extended to include PQC in parallel with QKD, combining both technologies simultaneously to augment the strength of each link.  Using the SDN paradigm, Madrid and Munich were connected through emulated QKD links to demonstrate the security transparency also for long-distance links.

Additionally, an E2E key exchange based on the application layer was proposed.  This method is very generic and easy to adapt to any infrastructure.  It does not require any specific type of network design but needs local management of key-forwarding requests.

Diversifying the key generation across disjoint network paths, e.g., via fiber, satellite, or mobile network links, adds extra security.  Moreover, if the key generation over such paths is based on different technologies, such as PQC and prospective long-distance QKD, this effect will be enhanced.  Multiple, different key-combination variants based on such multi-path and -technology key streams were realized in the presented connection of the testbeds.  This approach would also lead to a significant reduction of side channels that come about in real-world implementations of ideal protocols.

The findings presented in this work open the door to long-haul interconnectivity between QKD metro networks.  Multiple, quantum-safe technologies were combined in parallel to define the next generation of highly-secure, pan-European interconnectivity.  The proposal can readily be adopted and offers a scalable security layer that is extendable to all the continent and even across continents.

\funding{We would like to thank the project MadQ-CM (Madrid Quantum,  Comunidad de Madrid) funded by the European Union (NextGenerationEU, PRTR-C17.I1) and by the Comunidad de Madrid (Programa de Acciones Complementarias), the QUBIP project, funded by the European Union under the Horizon Europe framework program under Grant Agreement No. 101119746, QuantERA II Programme, European Union’s Horizon 2020 research and innovation program under Grant Agreement No 101017733, and with funding organizations: The Foundation for Science and Technology – FCT, Agence Nationale de la Recherche - ANR, and Spanish Agencia Estatal de Investigación – AEI, and the European Union’s Horizon Europe research and innovation funding program under the project ``Quantum Secure Networks Partnership'' (QSNP, grant agreement No 101114043).}

\begin{adjustwidth}{-\extralength}{0cm}

\reftitle{References}



\bibliography{biblio} 

\PublishersNote{}
\end{adjustwidth}
\end{document}